\documentstyle[12pt]{article}

\def\thefootnote{\fnsymbol{footnote}}

\newcommand{\eq}{\begin{equation}}
\newcommand{\en}{\end{equation}}
\newcommand{\eqa}{\begin{eqnarray}}
\newcommand{\ena}{\end{eqnarray}}

\newcommand{\PR}[1]{Phys.\ Rev.\ {\bf #1}}
\newcommand{\PRL}[1]{Phys.\ Rev.\ Lett.\ {\bf #1}}

\begin{document}
\begin{titlepage}
\vskip0.5cm
\begin{flushright}
DFTT 60/96\\
\end{flushright}
\vskip0.5cm
\begin{center}
{\Large\bf 
 A new classification scheme for Random Matrix Theories}
\end{center}
\vskip 1.3cm
\centerline{
M. Caselle\footnote{e--mail: caselle~@to.infn.it}}
\vskip 1.0cm
\centerline{\sl   Dipartimento di Fisica 
Teorica dell'Universit\`a di Torino}
\centerline{\sl Istituto Nazionale di Fisica Nucleare, Sezione di Torino}
\centerline{\sl via P.Giuria 1, I-10125 Torino, Italy}
\vskip 1.cm

\begin{abstract}
In the last few years several new Random Matrix Models have been proposed and
studied.
They have found application in various different contexts, 
ranging from the physics of 
mesoscopic systems to the chiral transition in lattice gauge theory. 
These new ensembles can be classified in terms of the same Dynkin diagrams
and root lattices  which are used in the classification of the Lie algebras.
\end{abstract}
\end{titlepage}

\setcounter{footnote}{0}
\def\thefootnote{\arabic{footnote}}

In the last few years  several new Random Matrix Theories (RMT) 
have been discussed in the literature. They can  be considered as
generalizations of the six ensembles which were originally proposed by Wigner
and Dyson~\cite{wd}, and share with them several features. 
At the same time they are also characterized  by some new properties, which make
them particularly apt to describe universal phenomena in a wide set of new
physical contexts ranging from the physics of mesoscopic systems~\cite{alw}
(where most 
of them were originally introduced and discussed) to lattice gauge 
theory~\cite{verb}.

It is the aim of this letter to organize these new ensembles within a 
compact classification scheme. As we shall see, several entries in our
classification will be left empty. They represent new RMT's which can be easily
constructed by using the techniques which will be discussed below but which have
not found yet  application in any physical context.

At the same time much progress has been done in the 
physics of quantum electronic transport in disordered wires~\cite{alw}. 
Several features of disordered wires  can be described by 
studying suitable ensembles of random transfer matrices. These ensembles are
completely different from the RMT's discussed above. The main difference is that
these transfer matrix ensembles are characterized by an additional 
``time'' coordinate and one is usually
interested in the time evolution of the eigenvalues distribution,
which is described by a suitable  Fokker-Planck (FP)  equation.
In the application to the disordered wires the role of ``time'' coordinate is
played by the length of the wire, the Fokker-Planck  equation
is known as Dorokhov Mello Pereyra Kumar
(DMPK) equation~\cite{dmpk} and describes the evolution of the transmission 
eigenvalues (and hence of the conductance) as the  length of the wire
increases. 

A major result of our analysis is that we can insert also these, apparently
unrelated, transfer matrix ensembles in our classification scheme.

Our classification is strongly related to that of the
irreducible Symmetric Spaces (SS) and  makes use  
of the same Dynkin diagrams
and root lattices  which are used in the classification of the Lie algebras.
Some of these results were anticipated in~\cite{cas}. A similar approach was 
also
recently discussed in~\cite{az}.
\vskip 0.2cm
Let us start by listing some known results about symmetric spaces. A more
detailed account can be found for instance in~\cite{helg}.

Let us take two Lie groups $G$ and $K$ such that $K\subset G$ and let 
${\bf G}$ and ${\bf K}$ be the corresponding Lie algebras. Decompose 
${\bf G}$ as: ${\bf G}~=~{\bf K}~+~{\bf L}$, then the coset manifold
$X\equiv G/K$ is a symmetric space if the following commutation relations hold:

\eq
[{\bf K},{\bf K}]\subset {\bf K}~,\hskip 0.5cm
[{\bf K},{\bf L}]\subset {\bf L}~,\hskip 0.5cm
[{\bf L},{\bf L}]\subset {\bf K}~.
\en

The exponential map from ${\bf G}$ to $G$  maps ${\bf L}$ into $X$.

There are three possibilities:

\begin{description}
\item{$X^+]$}~~
$G$ is semisimple and {\bf compact}, then $X$ has everywhere {\bf non-negative}
curvature.

\item{$X^-]$}~~
$G$ is semisimple and {\bf non-compact}, then $X$ has everywhere 
{\bf non-positive} curvature.

\item{$X^0]$}~~
$G$ is {\bf non-semisimple} and 
$[{\bf L},{\bf L}]=0$, then $X$ has everywhere 
{\bf zero} curvature.
\end{description}
\vskip 0.2cm

The following four properties of SS are particularly relevant for our
analysis:

\vskip 0.2cm
\begin{description}
\item{1]} {\sl  Triplicity.}
\vskip 0.1cm
SS always appear in triplets: $\{X^+,X^-,X^0\}$ 
of non-negative, non-positive and
zero curvature, with 
 the same subgroup $K$.
If ${\bf L}$ is the algebra which generates $X^-$, then $X^+$ is generated by
$i{\bf L}$ and $X^0$ coincides with ${\bf L}$.

\vskip 0.3cm
\item{2]} {\sl  Spherical coordinate system.}
\vskip 0.1cm
Each SS admits a {\sl spherical} coordinate systems, whose radial coordinates
can be obtained as the exponential map of the Cartan subalgebra of ${\bf L}$.
Notice however that  other  choices
of coordinates on the SS are also possible. These ``non-Cartan'' basis
become important if some extra symmetry is imposed to the system\cite{z}.

\vskip 0.3cm
\item{3]} {\sl  Complete classification.}
\vskip 0.1cm
Irreducible SS can be classified completely with techniques similar to those
used to classify the Lie algebras. 
A triplet of irreducible SS is  identified by its particular root 
lattice and by a set of positive integers $m_\alpha$ which are called ``root 
multiplicities''. 
 A ``root lattice'' is a lattice generated by  a  finite
set of non-zero vectors 
${\cal R}\equiv\{\alpha\}\subset {\cal V}$, where  ${\cal V}$ is a 
 $n$ dimensional vector space.
These vectors are called roots and must fulfil a set of very stringent
properties, which we shall not discuss here. The main consequence of these
constraints is that the root lattices can be classified completely.
They fall into five infinite classes which are
 conventionally denoted as $A_n$, $B_n$, $C_n$, $D_n$ and $BC_n$.  
The index $n$ denotes the rank of lattice which coincides with the dimensions of
${\cal V}$.
Exceptional isolated solution also exist for low values of $n$,
but they are not relevant for the present analysis.
The five infinite classes are defined as follows:
\begin{description}
\item{$A_n]$} 
As space ${\cal V}$ let us take the hyperplane in ${\bf R}^{n+1}$
for which $v_1+v_2+\cdots+v_{n+1}=0$, with $v_i\in {\bf R}^{n+1}$. 
Let us take a canonical basis in ${\bf R}^{n+1}$; $\{e_1,e_2,\cdots,e_{n+1}\}$.
Then ${\cal R}=\{e_i-e_j,~~i\neq j\}$

\item{$B_n$]} In this case let us take ${\cal V}={\bf R}^{n}$
Then ${\cal R}=\{\pm e_i,~\pm e_i\pm e_j,~~i\neq j\}$

\item{$C_n$]} ${\cal V}={\bf R}^{n}$ and
 ${\cal R}=\{\pm 2e_i,~\pm e_i\pm e_j,~~i\neq j\}$

\item{$D_n$]} ${\cal V}={\bf R}^{n}$ and
 ${\cal R}=\{\pm e_i\pm e_j,~~i\neq j\}$

\item{$BC_n$]} ${\cal V}={\bf R}^{n}$ and
 ${\cal R}=\{\pm e_i,\pm 2e_i,~\pm e_i\pm e_j,~~i\neq j\}$

\end{description}

The roots of type $\{\pm e_i\pm e_j,~~i\neq j\}$ are called ordinary roots
while $\{\pm 2e_i\}$ and $\{\pm e_i\}$ are called, respectively, long and short
roots. In the following we shall denote with $m_o$, $m_s$ and
$m_l$ the multiplicities of the ordinary, short and long roots
respectively.
 For each class of root lattices 
there are only few sets of $m_{\alpha}$ values which are
 compatible with the constraints which characterize the symmetric spaces. 
Altogether they generate 11 different infinite series of triplets, which, apart
from few exceptional solutions, exhaust the set of all possible irreducible
symmetric spaces.
They are listed in tab.I below.  To make more readable the table,
in the first column  only the coset description of the
$X^-$ element of each triplet  is reported and 
 the $SO(2p)/U(p)$ class has been
splitted into two separate entries.

\vskip 0.3cm
\item{4]} {\sl  Laplace Beltrami operator}
\vskip 0.1cm
The informations on the root lattice are enough to construct,
for each SS, the  explicit form of the radial part of the
Laplace-Beltrami operator $B$ in terms of the radial coordinates $\{q_i\}$:
\eq
B=\frac{1}{J^s(q)}\sum_{k=1}^n\frac{\partial}{\partial q_k}
J^s(q)\frac{\partial}{\partial q_k}
\label{fp}
\en
where $n$ is the rank of the SS, $s\in\{0,+,-\}$ and $J^s(q)$ is the Jacobian of
the transformation to the spherical coordinates and is defined as 
\eq
J^0(q)=\prod_{\alpha\in R^+} q_\alpha^{m_\alpha}\hskip 1cm
{\rm (zero~~curvature)}
\en
\eq
J^+(q)=\prod_{\alpha\in R^+} [sin(q_\alpha)]^{m_\alpha}\hskip 0.3cm
{\rm (positive~~curvature)}
\en
\eq
J^-(q)=\prod_{\alpha\in R^+} [sh(q_\alpha)]^{m_\alpha}\hskip 0.3cm
{\rm (negative~~curvature)}
\en
where $R^+$ is the subset of the positive roots of the lattice,
 $m_\alpha$ is the multiplicity of the root $\alpha$ and $q_\alpha$ denotes
the projection in $\{q\}$ of the root $\alpha$. E.g. if $\alpha=e_i-e_j$ then
$q_\alpha=q_i-q_j$.

\end{description}
\vskip 0.6cm

We are now in the position to list our main results. Some of them are already
well known in the literature, even if they are rephrased here in a different
language. They are collected here only for completeness. Some of them  however
are new and require some further comment.
\vskip 0.2cm
\begin{description}
\item{a1]} 
To each symmetric space it can be associated a suitable RMT by mapping
the radial coordinates of the SS in
the eigenvalues of the RMT.
The root multiplicities $m_\alpha$ 
correspond to the critical indices of the RMT.
Like the symmetric spaces, also the RMT's
can be organized in triplets: circular,
gaussian and transfer matrix ensembles with the same eigenvalue content, the
same symmetries and the same critical indices. 
Then the following relations hold:
\vskip 0.3cm
\begin{center}
SS of {\bf  positive} curvature correspond to {\bf circular} ensembles
\vskip 0.1cm

those of {\bf  negative} curvature to {\bf  transfer matrix} ensembles
\vskip 0.1cm

and those of {\bf  zero} curvature to {\bf  gaussian} ensembles.
\end{center}
\vskip 0.3cm

These RMT's are defined, in the case of the ensembles of gaussian or
circular type, by the following joint probability density for the
eigenvalues:
\eq
{\rm gaussian}\hskip 1cm P(x_i)=J^0(x) e^{-\sum_{j=1}^n x_j^2}
\en
\eq
{\rm circular}\hskip 1cm P(x_i)=J^+(x) 
\en
 and in the case of the transfer matrix ensembles  by the following
Fokker-Planck equation for the time dependent probability density
$P(x_i,t)$ :

\eq
f\frac{\partial P}{\partial t}=
\sum_{k=1}^n\frac{\partial}{\partial x_k}
[J^-\frac{\partial}{\partial x_k}\frac{1}{J^-}~P]
\equiv D ~P
\label{fp2}
\en
where the constant $f$ sets the units in which the time $t$ is measured.
Also for the gaussian and circular RMT it is possible to write FP
equations which  describe the approach to the equilibrium configuration of the 
ensembles. They have the same form of eq.(\ref{fp2}), but with $J^0$ or $J^+$
instead of $J^-$. One can easily see that
the differential operator $D$
 which appears in the Fokker-Planck eq. of the RMT is simply
related to
 the radial part $B$ of the Laplace-Beltrami operator on the SS:
\eq
D=J^s B~[J^s]^{-1}
\en
As a consequence, several properties of these FP
equations and in some cases also asymptotic solutions can be obtained by only
resorting to group theoretical arguments.

\vskip 0.2cm
\item{a2]} 
 A classification similar to that described above for the SS also holds
 for triplets of RMT's. However
if some extra symmetry imposes the choice of a
 non-Cartan basis then, following the same steps of point [a1] above one can
 construct new joint probability densities for the eigenvalues, whose critical
 indices could well lie outside the Cartan classification of Tab.I~\cite{z}.
We shall see below some realizations of this type
which have relevant physical applications.
According to this observation we shall keep in the following 
 the values of the root multiplicities
unconstrained. All the following results hold for generic
values of $m_\alpha$.

\vskip 0.2cm
\item{a3]} 
As we mentioned above, the root multiplicities $m_\alpha$ 
correspond to the critical indices of the RMT.
In particular the multiplicities of the long and short roots correspond to the
boundary critical indices while that of the ordinary roots corresponds to the
``bulk'' critical index of the RMT, the one which is usually denoted with
$\beta$.  
\vskip 0.2cm
\item{a4]} 
Some relevant symmetries of the RMT's can be understood as 
symmetries of the associated root lattice. In particular RMT's of $A_n$ type are
characterized by the translational invariance of the eigenvalues, while for all
the other lattices ($B_n$ $C_n$ $D_n$ and $BC_n$) this invariance is broken,
but a new $Z_2$ symmetry appears. The $A_n$ ensembles of positive or zero
curvature (namely of circular or gaussian type) exactly coincide with the
original Wigner-Dyson (WD) ensembles 
which, as it is well known, are the only ones
which fulfil such a translational invariance. All the other ensembles are
characterized by the presence of a boundary (even if sometimes it is not
immediately evident). They are not translational invariant, but 
are characterized by a new symmetry, the reflection
with respect to the boundary, which is exactly the $Z_2$ symmetry mentioned
above. In the following we shall refer to these ensembles, in opposition to the
WD ensembles, as boundary random matrix theories (BRMT)

\vskip 0.2cm
\item{a5]} 
To each  RMT of the circular or gaussian type,
 it is possible to associate a set of {\sl Classical} 
Orthogonal Polynomials, which
can be used  to construct correlations functions. 
As it is well known, the Hermite polynomials are related to 
the Wigner-Dyson ensembles ($A_n$ root
lattices). Simple changes of variables allows to show that
the Laguerre polynomials are related  to the gaussian BRMT and 
that the Jacobi ones to
the circular BRMT. In these last two cases the boundary 
critical indices  of the BRMT are
related to the parameter which define the polynomials as follows:
\vskip 0.1cm
{\sl Laguerre:}
\eq
L^{(\lambda)}\equiv \frac{x^{-\lambda}e^x}{n!}\frac{d^n}{dx^n}
(x^{n+\lambda} e^{-x}),\hskip 0.5cm (x\geq 0)
\en
\eq
\lambda=\frac{m_s+m_l-1}{2}
\en
\vskip 0.1cm
{\sl Jacobi:,\hskip 0.5cm $(-1\leq x\leq 1)$}
\eq
P^{(\mu,\nu)}\equiv \frac{(-1)^n}{2^nn!}
\frac{(1-x)^{-\nu}}{(1+x)^\mu}
\frac{d^n}{dx^n}
\{\frac{(1+x)^{n+\mu}}{(1-x)^{-n-\nu}} \}
\en
\eq
\mu=\frac{m_s+m_l-1}{2},\hskip 1cm \nu=\frac{m_l-1}{2}
\en

We see that $\lambda$ and $\mu$ have the same expression in terms of $m_s$ and
$m_l$. Thus the BRMT's corresponding to Laguerre and Jacobi polynomials with the
same $\lambda=\mu$ indices belong to the same triplet in the above
classification. They are respectively the
zero curvature (Laguerre) and positive curvature (Jacobi) elements of the
triplet. This explains the ``weak universality'' recently observed for the
boundary critical indices of these ensembles~\cite{for}.

\end{description}

\def\mycaptionl#1#2{
\begin{center}
\hskip 1pt\vskip -1cm
\begin{minipage}{14cm}
\small {\bf #1}: {\sl #2}
\end{minipage}
\null\hskip 1pt\vskip -0.2cm
\end{center}}

\begin{table}[ht]
\mycaptionl{\bf Table I}{
Irreducible Symmetric Spaces and some of their RMT realizations.} 
\label{tab1}
\hskip -1.cm
\begin{tabular}{|c|c|c|c|c|c|}
\hline
$ G/K$ & ${\cal R}$ & $m_\alpha$ & $X^+$ & $X^0$ & $X^-$  \\  
\hline
$SL(n,{\bf R})/SO(n)$ & $A_{n-1}$ & $m_o=1$ & {\rm COE} & {\rm GOE}& \\
\hline
$SL(n,{\bf C})/SU(n)$ & $A_{n-1}$ & $m_o=2$ & {\rm CUE} & {\rm GUE}& \\
\hline
$SU^*(2n)/Sp(2n)$ & $A_{n-1}$ & $m_o=4$ & {\rm CSE} & {\rm GSE}& \\
\hline
$\frac{SO(2n+1,{\bf C})}{SO(2n+1)}$
 & $B_{n}$ & $m_o=m_s=2$ &  & & \\
\hline
$SO(2n,{\bf C})/SO(2n)$ & $D_{n}$ & $m_o=2$ &  {\rm NS-D } & {\rm NS-D }& \\
\hline
$Sp(2n,{\bf C})/Sp(2n)$ & $C_{n}$ & $m_o=m_l=2$ &  {\rm NS-C } & {\rm NS-C }& \\
\hline
$Sp(2n,{\bf R})/U(n)$ & $C_{n}$ & $m_o=m_l=1$ & {\rm NS-CI } & {\rm NS-CI }& 
{\rm TOE}\\
\hline
$SO^*(4n)/U(2n)$ & 
 $C_n$& 
 $m_o=4,~m_l=1$ & {\rm NS-DIII} 
 & {\rm NS-DIII}&{\rm TSE} \\
\hline
$SO^*(4n+2)/U(2n+1)$ & 
 $BC_n$& 
 $m_0=m_s=4$,$m_l=1$& 
 & & \\
\hline
$\frac{SO(p,q)}{SO(p)\otimes SO(q)},~~(p\geq q)$ & $B_{q},D_q$
 & \vbox{\hbox{$m_o=1$}\hbox{$m_s=(p-q)$}}
 & {\rm } & {\rm chGOE}& \\
\hline
$\frac{SU(p,q)}{S(U(p)\otimes U(q))},~~(p\geq q)$ & $BC_{q}$
 & \vbox{\hbox{$m_o=2$}\hbox{$m_s=2(p-q)$}\hbox{$m_l=1$}}
 & {\rm (p=q)~ SUE} &  {\rm chGUE}&  {\rm (p=q)~TUE} \\
\hline
$\frac{Sp(2p,2q)}{Sp(2p)\otimes Sp(2q)},~~(p\geq q)$ & $BC_{q}$
 & \vbox{\hbox{$m_o=4$}\hbox{$m_s=4(p-q)$}\hbox{$m_l=3$}}
 &  &  {\rm chGSE}& {\rm (p=q)~\cite{cas3}}\\
 \hline

\end{tabular}
\end{table}

Let us conclude by listing some known examples of RMT's and giving their
collocation in the classification scheme.

\vskip 0.3cm
\begin{description}
\item{\sl Wigner-Dyson ensembles}
\vskip 0.1cm
These ensembles correspond to the $A_n$ type SS discussed above. In this case we
have only ordinary roots, and $J^0$ and $J^+$ become:
\eq
J^0(\{x_n\})=\prod_{i<j}~|x_j-x_i|^{m_0}
\en
\eq
J^+(\{\theta_n\})=\prod_{i<j}~|e^{i\theta_j}-e^{i\theta_j}|^{m_0}
\en
taking into account that for the three SS of the $A_n$ type we have
 $m_o=1,2,4$, we immediately
recognize the joint probability distribution of the three circular and gaussian
WD ensembles.

\vskip 0.3cm
\item{\sl Chiral RMT's}
\vskip 0.1cm
These ensembles correspond to the three last rows of table I. They are of
gaussian type and their joint probability distribution is given by:
\eq
J^0(\{x_n\})=\prod_{i<j}~|x_j-x_i|^{m_0}~\prod_jx_j^{\frac{m_s+m_l-1}{2}}~.
\en
They are also known as Laguerre ensembles and 
 find applications in several physical contexts, ranging from quantum 
wires~\cite{ns} to lattice QCD~\cite{verb}.

\vskip 0.3cm
\item{\sl Transfer matrix ensembles: Disordered wires}
\vskip 0.1cm
These ensembles are related to symmetric spaces of negative curvature.
Thus they are better defined in terms of their Fokker-Planck evolution
equation. Simple algebraic manipulations show that in the three cases
corresponding to the TOE, TUE and TSE entries of tab.I , $J^-$ is given by:
\eq
\hskip -0.3cm J^-(\{x_n\})=\prod_{i<j}~|sh^2x_j-sh^2x_i|^{\frac{m_0}{2}}
~\prod_j|sh~2x_j|^{\frac{m_l}{2}}
\en
with $m_o=1,2,4$ and always $m_l=1$.
If inserted in eq.(\ref{fp2}) this expression 
leads to the well known DMPK equation~\cite{dmpk}.
Recently, a generalization with $m_l=3$ has also been proposed~\cite{cas3}.

\vskip 0.3cm
\item{\sl NS ensembles}
\vskip 0.1cm
These are four new RMT's of gaussian and circular type which have been recently 
discussed in~\cite{az}.
They  describe the behaviour of hybrid microstructure composed of
normal metallic conductors in contact with superconducting regions. 
They are reported in tab.1 keeping the notation $\{C,CI,D,DIII\}$ used
in~\cite{az}.

\vskip 0.3cm
\item{\sl Ballistic chaotic quantum dots}
\vskip 0.1cm
These are three ensembles of the circular type, which have been introduced
in~\cite{sm} 
to describe the conductance of ballistic quantum dots, (characterized
by a chaotic classical dynamics). We shall denote them in the following as
S matrix ensembles.
They are described by root lattices of the $BC_n$ type with the following root
multiplicities: $m_o=\beta,~m_s=\beta-2,~m_l=1$, where $\beta$ can take, as 
usual the three values $\beta=1,2,4$. These three values exhaust the three
possible ensembles of this type  
which are  denoted, with abuse of language, as
orthogonal, unitary and symplectic:  SOE, SUE and SSE in short.

This last example is probably the most interesting, because only one of the
three ensembles corresponds to the Cartan decomposition of a symmetric space. 
It is the unitary
($\beta=2$) case (which is actually described by a $C_n$ lattice, since for
$\beta=2$, $m_s=0$). The other two cases correspond to basis of non-Cartan 
type (they can be realized as circular $A_{2n}$ ensembles with the imposition of
the  extra
symmetry $U(n)\times U(n)$~\cite{z,sm}), accordingly  
the root
multiplicities lie outside the classification of tab.I.

\end{description}
\vskip 0.5cm
{\bf Acknowledgements}

I wish to thank F.Gliozzi and M.Zirnbauer for a careful reading of the preprint
and several useful suggestions.

\end{document}